\documentclass[aps, prd, preprint, superscriptaddress, showpacs]{revtex4}
\usepackage{graphicx,epstopdf}
\usepackage{ulem}
\usepackage{epsfig}
\usepackage{color}
\usepackage{multirow}
\usepackage{epstopdf}
\definecolor{My_red}        {cmyk}{0.00, 1.00, 1.00, 0.20}

\newcommand{\bmat}{\left(\begin{array}}
\newcommand{\emat}{\end{array}\right)}
\newcommand{\beq}{\begin{equation}}
\newcommand{\eeq}{\end{equation}}

\newcommand{\lsim}{\mathrel{\ltap}}

\usepackage[centertags]{amsmath}
\usepackage{amssymb}

\newcommand{\missingET}{\mathchoice{\rlap{\kern.2em/}E_T}{\rlap{\kern.2em/}E_T}{\rlap{\kern.1em$\scriptstyle/$}E_T}{\rlap{\kern.1em$\scriptscriptstyle/$}E_T}}

\let\jnfont=\rm
\def\NPB#1,{{\jnfont Nucl.\ Phys.\ B }{\bf #1},}
\def\PLB#1,{{\jnfont Phys.\ Lett.\ B }{\bf #1},}
\def\EPJC#1,{{\jnfont Eur.\ Phys.\ Jour.\ C }{\bf #1},}
\def\PRD#1,{{\jnfont Phys.\ Rev.\ D }{\bf #1},}
\def\PRL#1,{{\jnfont Phys.\ Rev.\ Lett.\ }{\bf #1},}
\def\MPLA#1,{{\jnfont Mod.\ Phys.\ Lett.\ A }{\bf #1},}
\def\JPG#1,{{\jnfont J.\ Phys.\ G }{\bf #1},}
\def\CTP#1,{{\jnfont Commun.\ Theor.\ Phys.\ }{\bf #1},}
\def\JHEP#1,{{\jnfont JHEP \ }{\bf #1},}
\def\NPPS#1,{{\jnfont Nucl.\ Phys.\ Proc.\ Suppl.\ }{\bf #1},}
\def\CPC#1,{{\jnfont Computl.\ Phys.\ Commun.\ }{\bf #1},}
\def\CPL#1,{{\jnfont Chin.\ Phys.\ Lett. }{\bf #1},}
\def\APPB#1,{{\jnfont Acta\ Phys.\ Polon.\ B }{\bf #1},}

\def\lsim{\raise0.3ex\hbox{$<$\kern-0.75em\raise-1.1ex\hbox{$\sim$}}}
\def\gsim{\raise0.3ex\hbox{$>$\kern-0.75em\raise-1.1ex\hbox{$\sim$}}}

\begin{document}
\preprint{TTP14-010}

\title{Pseudo-goldstino and electroweak gauginos at the LHC }

\author{Ken-ichi Hikasa}
\affiliation{Department of Physics, Tohoku University, Sendai 980-8578, Japan}

\author{Tao Liu}
\affiliation{Institut f\"ur Theoretische Teilchenphysik, Karlsruhe
  Institute of Technology (KIT), D-76128 Karlsruhe, Germany}

\author{Lin Wang}
\affiliation{Institut f\"ur Theoretische Teilchenphysik, Karlsruhe
  Institute of Technology (KIT), D-76128 Karlsruhe, Germany}

\author{Jin Min Yang}
\affiliation{State Key Laboratory of Theoretical Physics,
      Institute of Theoretical Physics, Academia Sinica, Beijing 100190,
      China \vspace{2cm}}

\begin{abstract}
The multi-sector SUSY breaking predicts the existence of pseudo-goldstino,
which could couple more strongly to visible fields than ordinary
gravitino. Then the lightest neutralino and chargino can decay into
a pseudo-goldstino plus a $Z$-boson, Higgs boson or $W$-boson. In this note
we perform a Monte Carlo simulation for the direct productions of
the lightest neutralino and chargino followed by the decays to
pseudo-goldstino. Considering scenarios with higgsino-like, bino-like or
wino-like lightest neutralino, we find that the signal-to-background
ratio at the high luminosity LHC is between 6 and 25\%\ and the
statistical significance can be above $5\sigma$.
\end{abstract}

\maketitle

\section{Introduction}
Supersymmetry (SUSY) remains the most popular theory for solving the
hierarchy problem, albeit the recent discovery of a 125 GeV Higgs
boson, which makes most low energy SUSY models suffer from
fine-tuning to some extent \cite{susy-status}.  From the viewpoint of
model-building, the mechanism of SUSY breaking remains a puzzle.
Usually, it is assumed that spontaneous breaking of SUSY occurs
in some hidden sector and is mediated to visible fields by certain
mechanism. Then a massless fermion named goldstino appears,
which in the existence of local SUSY is absorbed into the longitudinal
component of gravitino. If
SUSY is broken in multiple sectors independently, each sector gives
a goldstino $\eta_i$ with SUSY breaking scale $F_i$. One linear
combination of $\eta_i$ is massless and eaten by the gravitino,
while the orthogonal combination remains as a physical state and is
named pseudo-goldstino. The property and related phenomenology of pseudo-goldstino
have been investigated in the literature
\cite{Cheung:2010mc,Cheung:2010qf,Benakli:2007zza,Craig:2010yf,McCullough:2010wf,Izawa:2011hi,Thaler:2011me,Cheung:2011jq,Bertolini:2011tw,Cheng:2010mw,Argurio:2011hs,Mawatari:2011jy,Argurio:2011gu,Ferretti:2013wya,Liu:2013sx}.
Comparing to the gravitino, the interactions of pseudo-goldstino are not totally
constrained by the supercurrent and thus some of its coulpings could
be large enough to have intriguing phenomenology. In the framework
of gauge mediated SUSY breaking (GMSB), pseudo-goldstino can make final
states softer and more structured at colliders
\cite{Argurio:2011gu}. In GMSB with more
than two hidden sectors the multi-photon signature
was discussed in \cite{Ferretti:2013wya} and
the LHC detectability for the Higgs boson decay into a pseudo-goldstino
was examined in \cite{Liu:2013sx}.

The non-observation of sparticles at the 7 TeV and 8 TeV runs of the
LHC has set stringent bounds on colored sparticles. However, the
electroweak sparticles are less constrained because of their small
production rates, and can still have masses below 1 TeV.
Theoretically, a light spectrum of electroweak sparticles is
naturally predicted in some frameworks like anomaly mediation and
non-minimal gauge mediations. So the study of electroweak
sparticles, especially the light neutralinos and charginos, is
rather important for testing SUSY at the LHC. At the LHC the
neutralinos and charginos can be directly produced through the
Drell-Yan process and vector boson fusion. In many conventional
scenarios with $R$-parity, the lightest neutralino is stable and
just leads to missing energy in the experiments. But in some low
scale gauge mediation scenarios the lightest neutralino can decay
into a photon plus a gravitino. In the scenario of SUSY breaking in
two hidden sectors, the lightest neutralino can decay to a pseudo-goldstino
plus a $Z$-boson or Higgs boson. In this work we focus on such a
two-sector SUSY breaking scenario to study the LHC detectability for
the productions of lightest neutralino and chargino.

This work is organized as follows.
In Section II we will make a brief review on the framework with pseudo-goldstino
and discuss its possible effect on the neutralino and chargino decays.
Then in Section III we take an effective way to study the corresponding signal
at the LHC. Finally, we give our conclusions in Section IV.

\section{Theoretical Review}
Due to the non-renormalization theorem of superpotential, the
spontaneous SUSY breaking is communicated to visible fields through
the non-trivial K\"ahler potential $K$ and gauge kinetic function
$f$. After integrating the hidden sector fields and parameterizing
their information in a non-linear way \cite{Samuel:1982uh}
\begin{eqnarray}
X_i=\frac{\eta_i^2}{2F_i}+\sqrt{2}\theta\eta_i+\theta^2F_i,
\end{eqnarray}
the following representative term which contributes to the soft mass can be obtained
\begin{eqnarray}
K&=&\Phi^\dagger \Phi \sum_i\frac{m_{\phi,i}^2}{F_i^{2}}X_i^\dagger X_i,\\
f_{ab}&=&\frac{1}{g_a^2}\delta_{ab}\left(1+\sum_i\frac{2m_{a,i}}{F_i}X_i\right).
\end{eqnarray}
In the above equations, $\eta_i$ is the so-called goldstino and
$m_{\phi,a}$ are respectively the soft masses for the chiral fields
and gauginos. The trilinear $A$ terms and bilinear $B_\mu$ could
also be constructed easily and we do not list them for simplicity.
In the two-hidden-sector scenario with the definition
$F=\sqrt{F_1^2+F_2^2}$ and $\tan\theta=F_2/F_1$, the combination
$G=\eta_1\cos\theta+\eta_2\sin\theta$ is eaten by the super-Higgs
mechanism, while one pseudo-goldstino
$G^\prime=-\eta_1\sin\theta+\eta_2\cos\theta$ is left. After
substituting the expression of $X_i$ and making some rotations, we
get the interaction Lagrangian up to order $1/F_i$:
\begin{eqnarray}
\mathcal{L}_G&=&\frac{m_\phi^2}{F}G\psi\phi^*
-\frac{im_a}{\sqrt{2}F}G\sigma^{\mu\nu}\lambda^aF^a_{\mu\nu}+
\frac{m_a}{F}G\lambda^aD^a  \label{LG},\\
\mathcal{L}_{G^\prime}&=&\frac{\widetilde{m}_\phi^2}{F}G^\prime\psi\phi^*
-\frac{i\widetilde{m}_a}{\sqrt{2}F}G^\prime\sigma^{\mu\nu}\lambda^aF^a_{\mu\nu}+
\frac{\widetilde{m}_a}{F}G^\prime\lambda^aD^a .\label{Leta}
\end{eqnarray}
Here the parameters $m$ and $\widetilde{m}$ are defined as
\begin{eqnarray}
&& m_{a}=m_{a,1}+m_{a,2}, \quad
\widetilde{m}_{a}=-m_{a,1}\tan\theta+m_{a,2}\cot\theta, \nonumber \\
&& m^2_{\phi}=m^2_{\phi,1}+m^2_{\phi,2}, \quad
\widetilde{m}^2_{\phi}=-m^2_{\phi,1} \tan\theta+m^2_{\phi,2} \cot\theta,
\label{eq6}
\end{eqnarray}
In our analysis we assume a large hierarchy between
$F_1$ and $F_2$ (we assume $F_1\gg F_2$ so that $\cot\theta$ is very large).
In this case the pseudo-goldstino can couple more strongly to visible fields
than ordinary goldstino (gravitino). Further, we will consider a small $\tilde{m}_a$
which happens for a large $\cot\theta$ ($m_{a,1}\tan\theta$ is suppressed) and
a very small $m_{a,2}$ (such a tiny gaugino mass is easily achieved if the
SUSY breaking sector $F_2$ approximately preserves R-symmetry \cite{Komargodski:2009jf}).
In this special case, the pseudo-goldstino couplings with the photon or transverse
$Z$-boson, which are proportional to $\tilde{m}_a$ in Eq.(\ref{Leta}), are suppressed.
So in our following analysis we neglect
the pseudo-goldstino couplings with the photon or transverse
$Z$-boson.

Of course, a pseudo-goldstino should have a mass.
At tree level its mass comes from the intrinsic property of SUGRA.
Also it can get loop corrections, which are very model-dependent.
In our analysis we assume that the pseudo-goldstino is rather light
so that a neutralino can decay into a pseudo-goldstino plus a $Z$-boson.
So our numerical results are only applicable to a rather light pseudo-goldstino
(for a rather light pseudo-goldstino, say below 10 GeV, we can approximately
neglect its mass in numerical calculations).
The phenomenology of a rather massive pseudo-goldstino
was considered in \cite{Argurio:2011gu}.

Now we look at the effects of pseudo-goldstino in concrete models.
In the minimal supersymmetric standard model (MSSM), the Lagrangian
for the neutralinos and charginos is given by
\begin{eqnarray}
\mathcal
{L}=-\dfrac{1}{2}Y_{ij}\chi_{i}\chi_{j}h^{0} +
G_{ij}\chi_{i}^{\dagger}\bar{\sigma}^{\mu}\chi_{j}Z_{\mu}+ (I_{ij}\chi^{\dagger}_{i}\bar{\sigma}^{\mu}\chi^{+}_{j}+
L_{ij}\chi_{j}^{-\dagger}\bar{\sigma}^{\mu}\chi_{i})W_{\mu}^{-}+h.c.
\end{eqnarray}
Here $\chi_{i,j}$ represent the four neutralinos in the gauge eigenbasis
$\{\tilde{B},\widetilde{W}^0,\tilde{H}_{d}^0,\tilde{H}_{u}^0\}$ and their mass matrix
is given by
\begin{equation}
M_{\tilde{N}}=
\left(
\begin{array}{cccc}
M_1&0&-c_\beta s_W m_Z&s_\beta s_W m_Z \\
0&M_2&c_\beta c_W m_Z&-s_\beta c_W m_Z \\
-c_\beta s_W m_Z&c_\beta c_W m_Z&0&-\mu\\
s_\beta s_W m_Z&-s_\beta c_W m_Z&-\mu&0
\end{array}
\right).
\label{eq:mN}
\end{equation}
$\chi^{\pm}_{i,j}$ are charginos in the gauge eigenbasis
$\{\widetilde{W}^{+},\tilde{H}_u^+,\widetilde{W}^{-},\tilde{H}_d^- \}$ and their mass matrix
is given by
\begin{equation}
M_{\tilde{C}}=
\left(
\begin{array}{cc}
0&X^T \\
X &0
\end{array}
\right),\ \ \
 \ \ \
X=
\left(
\begin{array}{cc}
M_2&\sqrt{2}s_\beta m_W \\
\sqrt{2} c_\beta m_W &\mu
\end{array}
\right).
\label{eq:mC}
\end{equation}
The couplings to the physical Higgs and gauge bosons are given by
\begin{equation}
Y = \frac{1}{2} \left(
\begin{array}{c c c c}
0 & 0 & g' s_\alpha & g' c_\alpha \\
0 & 0 & - g s_\alpha & - g c_\alpha \\
g' s_\alpha & - g s_\alpha & 0 & 0 \\
g' c_\alpha & -g c_\alpha & 0 & 0
\end{array}\right),
\ \ \
G = \frac{g}{2 c_W}
\left(\begin{array}{c c c c}
0 & 0 & 0 & 0 \\
0 & 0 & 0 & 0 \\
0 & 0 & 1 & 0 \\
0 & 0 & 0 & -1
\end{array}\right),
\end{equation}
\begin{equation}
I = g
\left(\begin{array}{c c c c}
0 & 0 & 0 & 0 \\
-1 & 0 & 0 & 0 \\
0 & 0 & 0 & 0 \\
0 & \frac{1}{\sqrt{2}} & 0 & 0
\end{array}\right),\ \ \
L = g
\left(\begin{array}{c c c c}
0 & 0 & 0 & 0 \\
0 & 0 & 1 & 0 \\
0 & 0 & 0 & \frac{1}{\sqrt{2}} \\
0 & 0 & 0 & 0
\end{array}\right).
\end{equation}
Since the contribution in Eq.~(\ref{Leta}) is proportional to
 $\widetilde{m}^2_\phi/F$,
there are two pseudo-goldstino interaction terms  which should be added to
the above Lagrangian:
\begin{align}
y_{i}G^{\prime}\chi_{i}h^{0} +  \rho_{i}G^{\prime}\chi_{i}
\end{align}
with the parameters $y_i$ and $\rho_{i}$ given by
\begin{equation}
y = \frac{1}{\sqrt{2} F}  \left( \begin{array}{c} 0 \\ 0 \\ \widetilde{B}_\mu c_\alpha - \widetilde{m}_{H_d}^2 s_\alpha \\ \widetilde{m}_{H_u}^2 c_\alpha - \widetilde{B}_\mu s_\alpha \end{array} \right),
\ \ \
\rho  = \frac{v}{\sqrt{2} F} \left(\begin{array}{c} 0 \\ 0 \\ \widetilde{m}_{H_d}^2 c_\beta + \widetilde{B}_\mu s_\beta \\ \widetilde{m}_{H_u}^2 s_\beta + \widetilde{B}_\mu c_\beta \end{array} \right).
\end{equation}
In the above matrices,  $\alpha$ and $\beta$ are the mixing angles in the
Higgs sector with $\tan\beta=\langle H_u^0 \rangle /\langle H_d^0 \rangle$.
We used the notations $s_W=\sin\theta_W$, $c_W=\cos\theta_W$
($\theta_W$ is the Weinberg angle) and $s_\beta=\sin\beta$, $c_\beta=\cos\beta$.

\begin{figure}[htbp]
\includegraphics[scale=0.65]{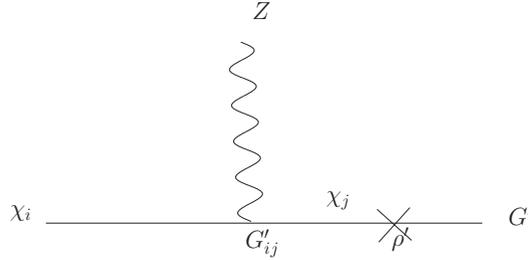}%
\vspace*{-0.5cm}
\caption{A diagrammatic show of interactions between $Z$-boson and pseudo-goldstino.}
\label{dia1}
\end{figure}

The linear terms induce a small mixing between neutralinos and
pseudo-goldstino, so we have to make a rotation to the mass eigenstate
basis for neutralinos and then the small mass mixing can be treated
perturbatively. For example, the vertex between $Z$-boson and
pseudo-goldstino $G^{\prime}$ appears after a mass insertion $\rho^\prime$,
as shown in Fig.~\ref{dia1}. The matrices $\rho_{i}^\prime$ and
$G_{ij}^\prime$ are defined as
\begin{align}
\rho_i^\prime = \rho_{j}N_{ji} , \quad  G_{ij}^\prime=  G_{\ell m}N_{\ell i}N_{mj}
\end{align}
where $N$ is the rotation to diagonalize the neutralino mass matrix.
Other interactions could be obtained in the same way, such
as the interaction between chargino and pseudo-goldstino. Since
in this scenario the couplings of pseudo-goldstino with photon
or transverse component of $Z$-boson are negligible,
the two possible decay channels for the lightest neutralino
are $Z$ or $h$ plus $G^\prime$.

From the above analysis we can get the structure of the interactions
for pseudo-goldstino. However, there are many parameters involved, especially
in the chargino and neutralino rotation matrices. So we only pick out
some representative interactions to study the corresponding phenomenology.

To study the phenomenology, we employ the effective Lagrangian
\begin{align}
\mathcal{L}_{\rm eff}=\dfrac{\widetilde m_\phi^2}{F}[g_{h\chi}h\chi^{0} G^{\prime}+g_{\chi Z}\bar{G^{\prime}}\bar{\sigma}^{\mu}\chi^{0} Z_{\mu}
+ g_{\chi W_{1}} \bar{G^{\prime}} \bar{\sigma}^{\mu} \chi^{+} W^{-}_{\mu} +
g_{\chi W_{2}} \bar{G^{\prime}} \bar{\sigma}^{\mu} \chi^{-} W^{+}_{\mu} +h.c.].
\label{eff}
\end{align}
Here we list all possible couplings, some of which may be turned off
in specific cases.
The decay widths of the lightest neutralino and chargino to pseudo-goldstino
are given by
\begin{eqnarray}
\Gamma(\chi^{0}\rightarrow h G^{\prime})&=&
\dfrac{m_{\chi}}{16\pi}\dfrac{g_{h\chi}^{2}\widetilde m_\phi^{4}}{F^{2}}
\left(1-\dfrac{m_{h}^{2}}{m_{\chi}^{2}}\right)^{2},\label{eq16}\\
\Gamma(\chi^{0}\rightarrow Z G^{\prime})&=&
\frac{m_{Z}^{2}}{8\pi m_{\chi}} \frac{g_{\chi Z}^{2}\widetilde m_\phi^{4}}{F^{2}}
\left(1-\frac{m_{Z}^{2}}{m_{\chi}^{2}}\right)
\left[\dfrac{m_{\chi}^{4}}{2m_{Z}^{4}}
+\dfrac{m_{\chi}^{2}}{2m_{Z}^{2}}-1\right],\label{eq17}\\
\Gamma(\chi^{\pm}\rightarrow W^{\pm} G^{\prime})&=&
\dfrac{m_{W}^{2}}{16\pi m_{\chi}} \dfrac{(g_{\chi W_{1}}^{2}
+g_{\chi W_{2}}^{2})\widetilde m_\phi^{4}}{F^{2}}
\left(1-\dfrac{m_{W}^{2}}{m_{\chi}^{2}}\right)
\left[\dfrac{m_{\chi}^{4}}{2m_{W}^{4}}+\dfrac{m_{\chi}^{2}}{2m_{W}^{2}}-1\right].
\label{eq18}
\end{eqnarray}
The first two decay modes have been considered in \cite{Thaler:2011me}
and we checked that our results agree with theirs.
In our calculation we fix $\widetilde m_\phi/\sqrt{F}=0.1$
and all the couplings $g_X$ to be unity. Under these assumptions, the weak
scale neutralino or chargino have the decay width at the
order of $\sim 10^{-4}$ GeV and the decay length
$\Gamma^{-1}\sqrt{(E^2-m_\chi^2)/m_\chi^2}\sim 10^{-10}$ cm so they
will decay inside the detector. Note that these parameters have no
effects on the production rates of neutralino or chargino.
As long as the neutralino and chargino only decay to pseudo-goldstino,
their signal rates  are not sensitive to these parameters.

About the parameter space in the neutralino/chargino sector,
following the analysis in \cite{Shufang Su}, we classify it
according to the relative values of $M_{1,2}$ and $\mu$:
 (i) $|\mu| < M_{1}, M_{2}$;
 (ii) $M_2 < M_{1}, |\mu|$;
 (iii) $M_1 < M_{2}, |\mu|$.
Each case corresponds to a different property of the lightest
neutralino, called the lightest ordinary sparticle (LOSP). In the
first case, the LOSP is higgsino-like, which can not only decay to
Higgs, but also decay to $Z$-boson though a mass insertion of
$\rho$. In the second and third cases the LOSP is respectively
wino-like and bino-like, which only decays to a Higgs boson plus a
goldstino through its mass mixing with the higgsino. For the
lightest chargino, which is too light to decay into a neutralino plus
an on-shell $W$-boson,
it now can decay into a $W$-boson plus a pseudo-goldstino. Note that in
the second case the interaction vertex needs
more than one insertion, so wino may mainly decay to gravitino.
Since the decay to gravitino has the same collider signature, we
assume the lightest chargino totally decay to pseudo-goldstino.

Note that in addition to the above decays,
the neutralino can also decay to a real goldstino (gravitino), which may
be competitive and need to be checked. The corresponding decay widths
are given by ~\cite{Ambrosanio:1996jn,Dimopoulos:1996yq}
\begin{eqnarray}
\Gamma(\chi^{0}\rightarrow \gamma G)&=&
|N_{11}c_{W}+N_{12}s_{W}|^2\dfrac{m_{\chi}^5}{16\pi F^2},\\
\Gamma(\chi^{0}\rightarrow Z G)&=& \left(|N_{11}s_{W}-N_{12}c_{W}|^2+
\frac{1}{2}|N_{13}c_{\beta}-N_{14}s_{\beta}|^2\right)\left(1-\dfrac{m_{Z}^2}{m_{\chi}^2}\right)^4
\dfrac{m_{\chi}^5}{16\pi F^2}.
\end{eqnarray}
Here we see that the decay $\chi^{0}\to \gamma G$ is suppressed for a higgsino-like
neutralino.
So in the following we demonstrate the results for a bino-like neutralino and
compare with the decays into a pseudo-goldstino.
For numerical calculations, we fix the parameters
$\tan\beta=10$, $M_{1}=200$~GeV, $M_{2}=500$~GeV and $\mu=1.0$~TeV.
The soft mass $\tilde m_\phi$ is a
combination of Higgs soft parameters whose values can be obtained from
SOFTSUSY \cite{Softsusy} once $\tan\beta$, $\mu$ and the SM-like Higgs
mass (we take 125 GeV) are fixed. Note that these Higgs soft parameters
receive contributions from two SUSY-breaking sectors and we assume the
two contributions are equal (say $B_{\mu,1}=B_{\mu,2}$) in the following
numerical example.

\begin{figure}[htbp]
\includegraphics[scale=0.85]{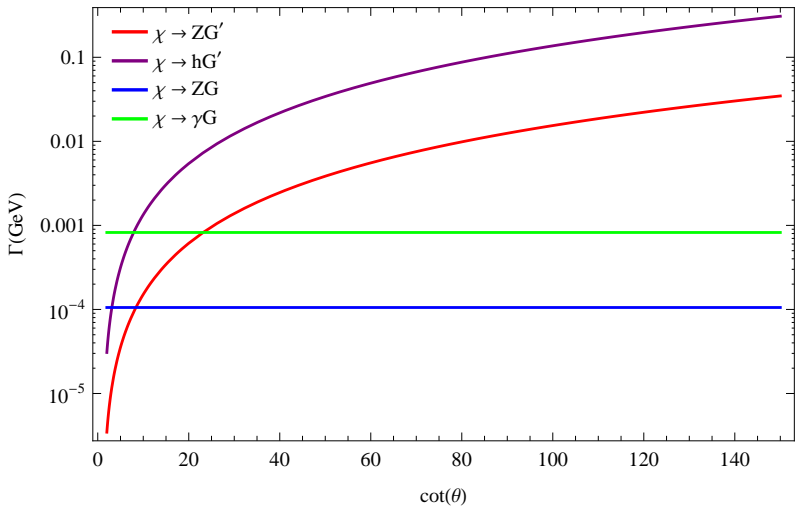}%
\vspace*{-0.5cm}
\caption{The partial widths of a bino-like neutralino, decaying to a pseudo-goldstino ($G^{\prime}$)
or real goldstino ($G$), as a function of $\cot\theta$. }
\label{width}
\end{figure}

With the above fixed parameters, we vary $\cot\theta$ and show the decay
widths in Fig.\ref{width}. As expected, for a small $\cot\theta$ the decays
into real goldstino are important while for a large $\cot\theta$ the decays
into pseudo-goldstino become dominant. The reason is obvious: the couplings of
pseudo-goldstino are proportional to $\tilde m_\phi$, which can be enhanced by
a large $\cot\theta$, as shown in Eq.(\ref{eq6}).

\section{Phenomenological study at LHC}
In this section we study the direct productions of the lightest
neutralino and chargino followed by the decays to pseudo-goldstino
at the LHC{}.
In our study we assume that other SUSY particles
(like squarks, sleptons, heavy Higgs bosons and gluino)
are heavy enough to be decoupled. The mass of the SM-like Higgs boson
is fixed at $m_{h} = 125$ GeV.
For the parameters $M_{1}$, $M_{2}$ and $\mu$,
they will be fixed with different values in three different cases
listed in the preceding section.
The sign of $\mu$ is assumed to be positive and $\tan\beta$ is fixed
as 10 in the calculation.
We use SOFTSUSY \cite{Softsusy} to calculate the mass spectrum and
the mixing matrices.

We use MadGraph5 \cite{MG5} to perform Monte Carlo simulations for
the signals and the SM backgrounds.
The effective Lagrangian in Eq.~(\ref{eff})
for the pseudo-goldstino interaction is implemented in FeynRules \cite{Feynrules}
and passed the UFO model file \cite{Ufo} to MadGraph5.
The signal and background samples are generated at parton
level by MadGraph5 and then passed to Pythia \cite{Pythia}
for parton shower and hadronization. The cross section of
the signal is normalized to the Next-to-Leading-Order (NLO) by using
Prospino2 \cite{prospino}. The fast detector simulations
are performed by using Delphes \cite{Delphes} with the ATLAS detector.
For the clustering jets we use the anti-$k_{t}$
algorithm \cite{anti-kt} with the radius parameter $\Delta R = 0.5$
in the FastJet package \cite{Fastjet}.
The sample analysis is performed with the package
MadAnalysis5 \cite{madanalysis5}.

\subsection{Higgsino-like LOSP ($|\mu| < M_{1}, M_{2}$)}
\begin{figure}[htbp]
\includegraphics[scale=0.5]{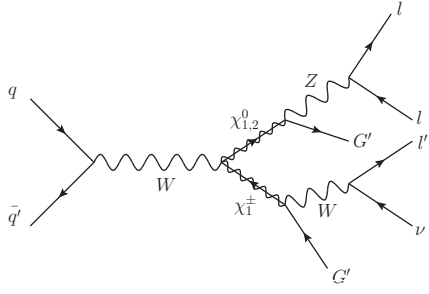}%
\caption{Feynman diagram for the pair production of
$\chi_{1,2}^{0}\chi_{1}^{\pm}$ followed by the subsequent decays
into pseudo-goldstino.} \label{diaFey}
\end{figure}

In this case the neutralino and chargino are produced mainly through the
pairs $\chi_{1}^{0}\chi_{1}^{\pm}$, $\chi_{2}^{0}\chi_{1}^{\pm}$,
$\chi_{1}^{+}\chi_{1}^{-}$, $\chi_{1}^{0}\chi_{2}^{0}$ (Note that if $\mu$ is much smaller
than $M_{1}$ and $M_{2}$, then the higgsino-like $\chi_{1}^{0}$,  $\chi_{2}^{0}$ and $\chi_{1}^{\pm}$
are nearly degenerate and such pair productions give no visible final states in the conventional
MSSM with $\chi_{1}^{0}$ being the LSP. In this case, to detect such productions at the LHC, an
extra jet or photon is needed \cite{mono-jet}).
Their cross sections
at the NLO can be found in \cite{Shufang Su}.
Among these channels the production of $\chi_{1,2}^{0}\chi_{1}^{\pm}$
has the largest rate.
In the two-hidden-sector SUSY breaking scenario, the neutralino
decays to a $Z$-boson or Higgs
plus a pseudo-goldstino $G^{\prime}$,  as discussed in Section II{}.
Due to the large systematic uncertainty for the Higgs hadronic
decay at the LHC, in this work we focus on the $Z$-boson mode
and assume its branching ratio to be $0.5$.
With the leptonic decays of $Z/W^{\pm}$, the signal is
\begin{equation}
pp\rightarrow\chi_{1,2}^{0}\chi_{1}^{\pm} \rightarrow Z G^{\prime} W^{\pm} G^{\prime} \rightarrow \ell^{+}\ell^{-}\ell^{\pm}\nu G^{\prime}G^{\prime}
\to 3\ell + \missingET 
,~~(\ell= e, \mu, \tau ).
\end{equation}
The relevant Feynman diagram is displayed in Fig.~\ref{diaFey}. Here
the three leptons in the final state contain an oppositely charged
lepton pair with same flavor. The tau lepton can be partially
reconstructed from its hadronic decays. Note that the neutralino
pair $\chi_{1}^{0}\chi_{2}^{0}$ can also contribute to the signal.
We checked that its contribution is very small and can be neglected
safely. The relevant mass parameters are fixed to $\mu=200$~GeV,
$M_{1}=1.0$~TeV and $M_{2}=1.5$~TeV as a benchmark scenario in the
calculation.

For the $3\ell+ \missingET$ final state, the dominant irreducible SM
background is the $WZ$ di-boson production. We also consider other
SM backgrounds including the top
quark pair production, the di-boson production of $ZZ$,
the $Z$-boson production in association with jets.
The top pair production with di-leptonic decays may fake the signal
since the $b$-jets and light jets may be misidentified as charged leptons.
The contribution from this process can be suppressed by applying
$b$-jets and light jets veto. For the background process $ZZ$
with both $Z$ bosons decaying to leptons, it can mimic our signal
when one of the leptons is missing in the detector.
In the case of $Z+j$ background, it may mimic our signal since a light jet
may fake to charged lepton. These processes could be suppressed by requiring
a large $ \missingET$.
We do not consider the multi-lepton ($n\geq$ 3) final state
from the production of three gauge bosons due to its small
cross section compared with other backgrounds.

\begin{figure}[htbp]
\includegraphics[scale=0.35]{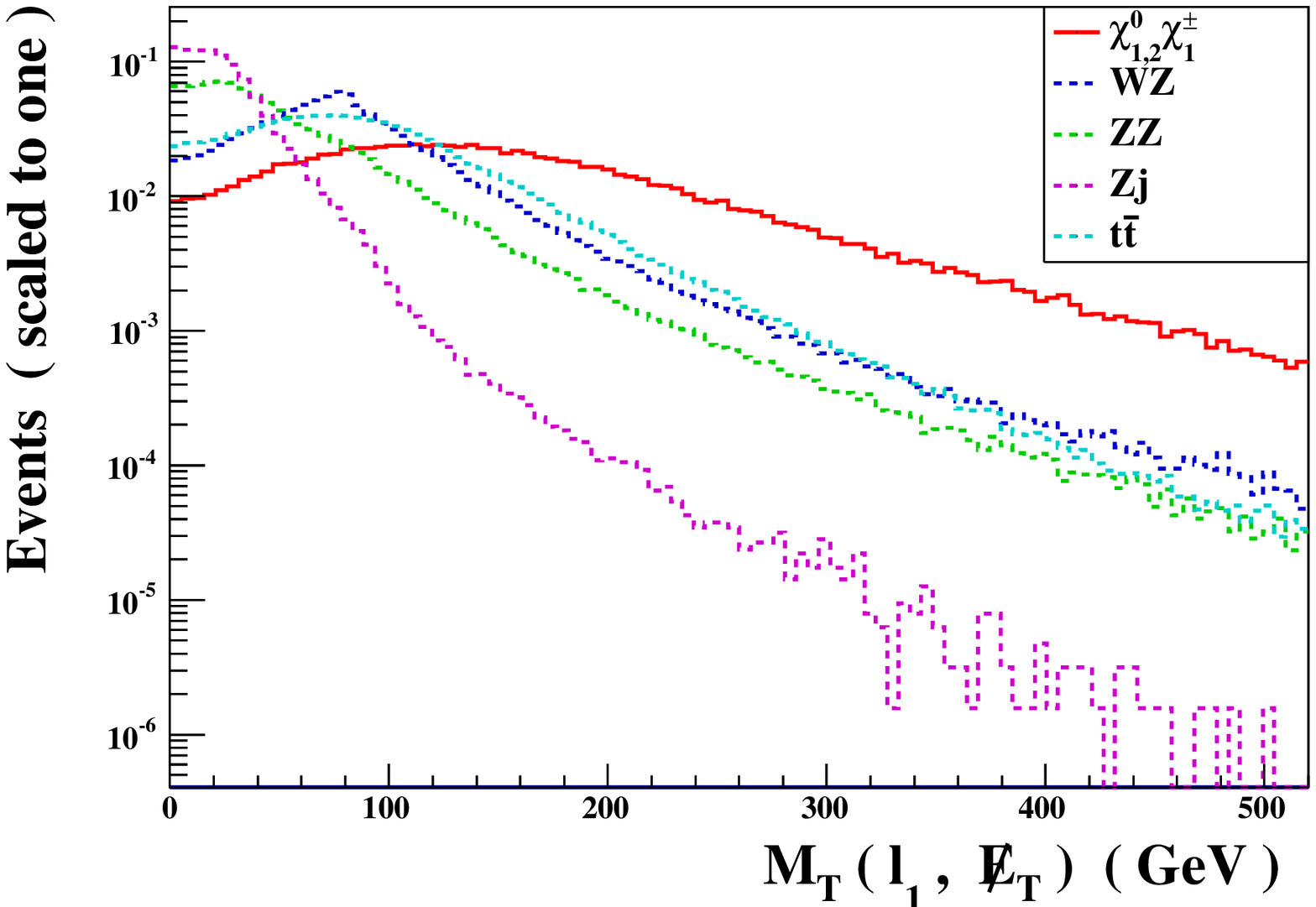}%
\includegraphics[scale=0.35]{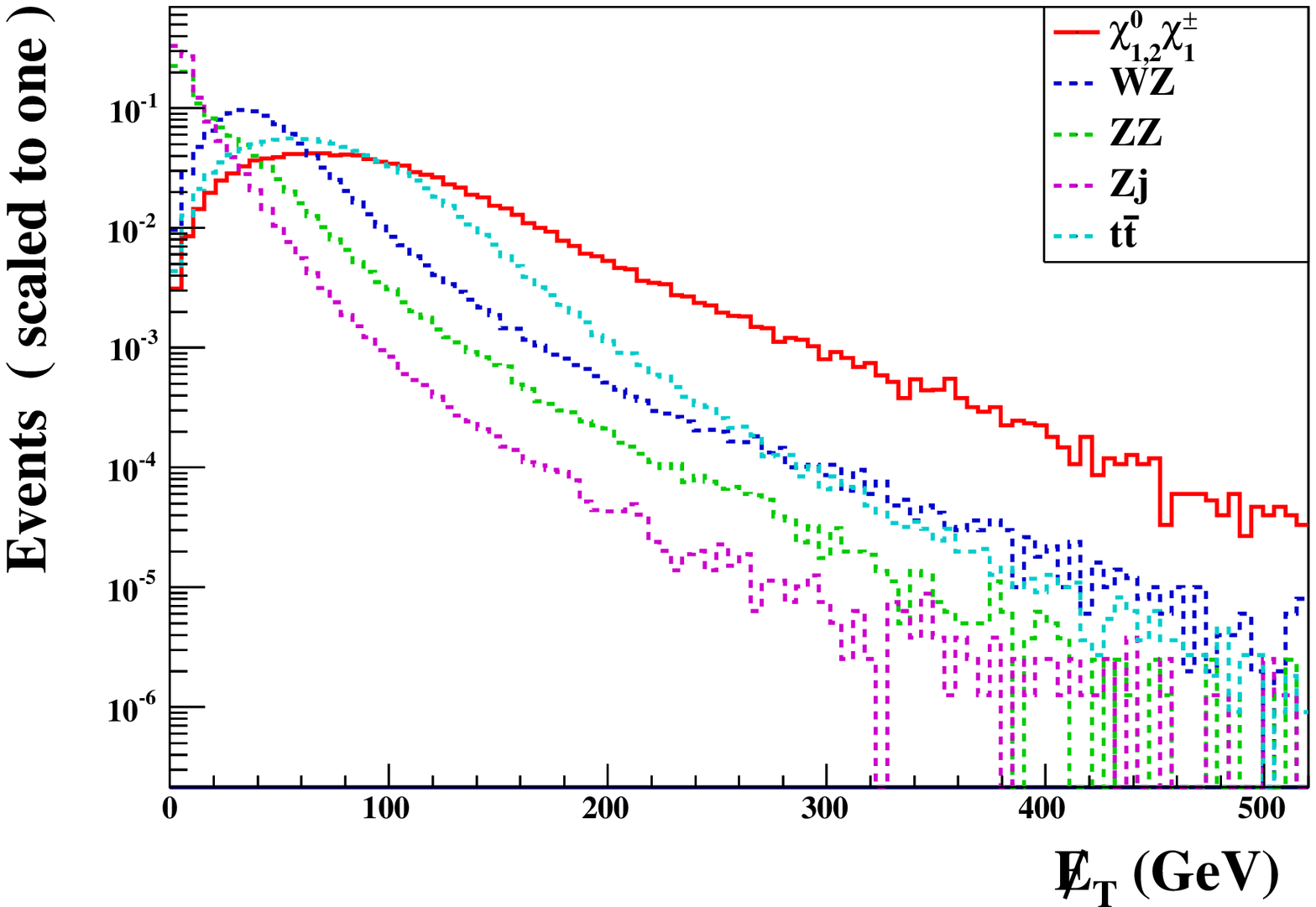}%
\hspace{0in}%
\caption{The normalized $M_{T}$ and $\missingET$ distributions
for the signal $pp\to\chi_{1,2}^{0}\chi_{1}^{\pm} \to Z G^{\prime} W^{\pm} G^{\prime}
\to \ell^{+}\ell^{-}\ell^{\pm}\nu G^{\prime}G^{\prime}
\to 3\ell + \missingET$
and background processes at the LHC with $\sqrt{s}=14$~TeV.
For the signal we fixed the relevant mass parameters
as $\mu=200$~GeV, $M_{1}=1.0$~TeV, $M_{2}=1.5$~TeV.}
\label{figetmt}
\end{figure}

To efficiently cut the SM backgrounds, we in Fig.~\ref{figetmt} plot
some kinematic distributions for the signal and the backgrounds at the LHC with
$\sqrt{s}=14$ TeV. In the left frame of Fig.~\ref{figetmt},
we give the normalized transverse mass $M_{T}(\ell_1,\missingET)$
distribution, where the definition of this variable is
\begin{equation}
M_{T}=\sqrt{2p_{T}^{\ell} \missingET [1 - \cos\Delta\phi_{\ell,\missingET}]},
\end{equation}
with $\Delta\phi_{\ell,\missingET}$ being the azimuthal angle
difference between the lepton and the missing energy.
Here we use the lepton with the largest transverse momentum
for constructing $M_T$. The right frame in Fig.~\ref{figetmt} shows the
normalized $\missingET$ distribution.
It is easy to see that a lower cut of about 120 GeV for $M_T$ and 100 GeV
for $\missingET$ can improve the statistical significance of
the signal. Based on these distributions, we apply the following event
selection:
\begin{itemize}
\item
Basic selection: three leptons with $p_{T}^{\ell_{1},\ell_{2},\ell_{3}} > 60,40,20$ GeV,
$\vert \eta \vert < 2.5$. We use the following isolation criterion for
electrons and muons: the transverse momentum sum of all charged particles
with $p_{T}^{\rm min} >0.5$ GeV that lie within a cone $R=0.5$ around electron or
muon should be less than $10\%$ of transverse momentum of central electron
or muon. Note that we assume the $\tau$-tagging efficiency to be $40\%$
and also include the mis-tags of QCD jets in Delphes.
\item
$M_{T}(\ell_1,\missingET) > 120$ GeV.
\item
$\missingET > 100$ GeV.
\item
The invariant mass of the oppositely charged lepton pair
with same flavor must be within $\vert m_{\ell\ell}-m_{Z} \vert < 20$ GeV.
\item
Veto on tagged $b$-jets with $p_{T}>20$ GeV and $\vert \eta \vert < 2.5$.
We use the $b$-jet tagging and $c$-jet mis-tagging efficiency parametrization
in \cite{b-tagging}. Delphes also includes misidentification rate for light
jets.
\item
Veto events with $p_{T}(j)>60$ GeV and $\vert \eta \vert < 5.0$.
\end{itemize}

In Table \ref{table1} we present the numbers of signal and
background events for the LHC with $\sqrt{s} = 14$ TeV and 100 fb$^{-1}$ of
integrated luminosity. We have normalized the cross section of the
$WZ$ production to NLO \cite{wznlo} and $t\bar{t}$ production to
next-to-next-to-leading order (NNLO) \cite{ttbnnlo}. From this table
we can see that the signal is overwhelmed by the backgrounds after
basic selection. As we excepted, the cut on the transverse mass
$M_{T}$ can suppress all the background processes significantly,
especially for the electroweak processes. They are further reduced
by requiring large missing transverse energy. Then the dominant
irreducible SM background $WZ$ is suppressed by about one order. The
large background $Zj$ has been completely removed. The other
important background $t\bar{t}$ is also reduced by about a factor of
seven. But the signal is decreased only a half. Though the invariant
mass of charged lepton pair cut $\vert m_{\ell\ell}-m_{Z} \vert <
20$ GeV reduces both the signal and backgrounds, it improves the
statistical significance of the signal efficiently. The final two
cuts vetoing on $b$-jets and light jets are of crucial importance to
further suppress the $t\bar{t}$ background.
Note that the veto on the light jet also has a small effect on the
signal due to the tau jet in the signal.
After all cuts, the signal-to-background ratio is 11\%.

\begin{table}
\caption{The numbers of events for signal
$pp\to\chi_{1,2}^{0}\chi_{1}^{\pm} \to Z G^{\prime} W^{\pm} G^{\prime}
\to \ell^{+}\ell^{-}\ell^{\pm}\nu G^{\prime}G^{\prime}
\to 3\ell + \missingET$
and backgrounds at the LHC with $\sqrt{s} = 14$ TeV and 100 fb$^{-1}$
of integrated luminosity. } \vspace{0.2cm}
\begin{tabular}{||c||c|c|c|c||c||}
 \hline \hline
\cline{1-6}
 cut &~~$W Z\rightarrow \ell\ell\ell\nu$~~  &  ~~$Z Z \rightarrow \ell\ell\ell\ell $~~  &  ~~$Z j \rightarrow \ell\ell j$~~  &  ~~$t \bar{t} \rightarrow bb\ell\ell\nu\nu$~~   & ~~signal~~ \\
 \hline
 basic selection &  7240  &   540  &  17133  &  24809  &  249  \\
 \hline
 $M_{T}(\ell_1,\missingET) > 120$ GeV  &  2690   &  86   &   1365   &   11824   &   205  \\
 \hline
 $\missingET > 100$ GeV  &  870  &  20  &  0   &  3563  &  129   \\
 \hline
 $|m_{\ell\ell}-m_{Z}|<20$ GeV   &  834  &  18  &  0   &  568  &  123   \\
 \hline
 veto on b-jets  &  832  &  18  &  0  &  438  &  123   \\
 \hline
 veto on light jet  &  781  &  15  &  0   &  237   &  114   \\
 \hline
 \hline
\end{tabular}
\label{table1}
\end{table}

\begin{table}
\caption{The numbers of signal events for
$pp\to\chi_{1,2}^{0}\chi_{1}^{\pm} \to Z G^{\prime} W^{\pm} G^{\prime}
\to \ell^{+}\ell^{-}\ell^{\pm}\nu G^{\prime}G^{\prime}
\to 3\ell + \missingET$
and its statistical significance
at the LHC with $\sqrt{s} = 14$ TeV and different luminosities.
$S_{1}$ and $B_{1}$ stand for the signal and background events after basic selection,
while $S_{2}$ and $B_{2}$ stand for the signal and
background events after all the cuts.} \vspace{0.2cm}
\begin{tabular}{||c|c|c|c|c|c|c||}
 \hline \hline
 ~~~$\sqrt{s} = 14$ TeV~~~  &  ~100 fb$^{-1}$~  &  ~200 fb$^{-1}$~  &  ~300 fb$^{-1}$~  &  ~400 fb$^{-1}$~  &  500 fb$^{-1}$  &  600 fb$^{-1}$  \\
\hline
 $S_{1{\rm [basic~selection]}}$  &  249  &  498  &  747  &  996  &  1245  &  1494  \\
 \hline
 $S_{2{\rm [passing~all~cuts]}}$ &  114  &  228  &  342  &  456  &  570  &  684  \\
  \hline
 $S_{1}$/$\sqrt{ S_{1}+B_{1}}$  &  1.1  &  1.6  &  1.9  &  2.2  &  2.5  &  2.7  \\
 \hline
 $S_{2}$/$\sqrt{ S_{2}+B_{2}}$  &  3.4  &  4.8  &  5.8  &  6.7  &  7.5  &  8.2  \\
 \hline
 \hline
\end{tabular}
\label{table2}
\end{table}

In Table \ref{table2} we show the number of signal events and its significance before and
after cuts for different luminosities at the 14 TeV LHC. Although the signal is reduced
by applying cuts, its statistical significance is increased efficiently.
With an integrated luminosity
of 200--300 fb$^{-1}$, the sensitivity can reach $5\sigma$.

\subsection{Wino-like LOSP ($M_{2} < M_{1}, \vert\mu\vert$)}
In this case, among the direct productions of neutralinos and
charginos at the LHC, the pair production of $\chi_{1}^{0}$
$\chi_{1}^{\pm}$ is dominant and we consider this process in our
analysis. As discussed before, the LOSP $\chi_{1}^{0}$ can only
decay to a Higgs boson and a pseudo-goldstino $G^{\prime}$ in this case.
Thus the signal is a single lepton and two bottom quarks with large
missing transverse energy:
\begin{equation}
pp\rightarrow\chi_{1}^{0}\chi_{1}^{\pm} \rightarrow h W^{\pm} G^{\prime}G^{\prime}
\rightarrow \ell^{\pm} b\bar{b} \nu G^{\prime}G^{\prime}
\to \ell+2b + \missingET
~~(\ell= e,\mu,\tau).
\end{equation}
In the calculation we fix the relevant parameters as $M_{2}=200$ GeV, $\mu=1.0$ TeV
and $M_{1}=1.5$ TeV. The other parameters are assumed to take the same values as in
the Higgsino-like case.

The dominant SM backgrounds for this signal are di-boson productions,
$Wb\bar{b}$, top pair and single top productions. For di-boson
productions, we only consider the $WZ$ production where $Z$ decays
to $b\bar{b}$ and $W$ decays leptonically. The contribution from
other di-boson productions should be very small. For the $Wb\bar{b}$
production, its contribution may be suppressed by requiring large
missing transverse energy. The top pair production can mimic the
signal if one of the $W$ bosons decays leptonically. The single top
production can also fake the signal when the light quark is
misidentified as a $b$-quark or missing transverse energy.

\begin{figure}[htbp]
\includegraphics[scale=0.34]{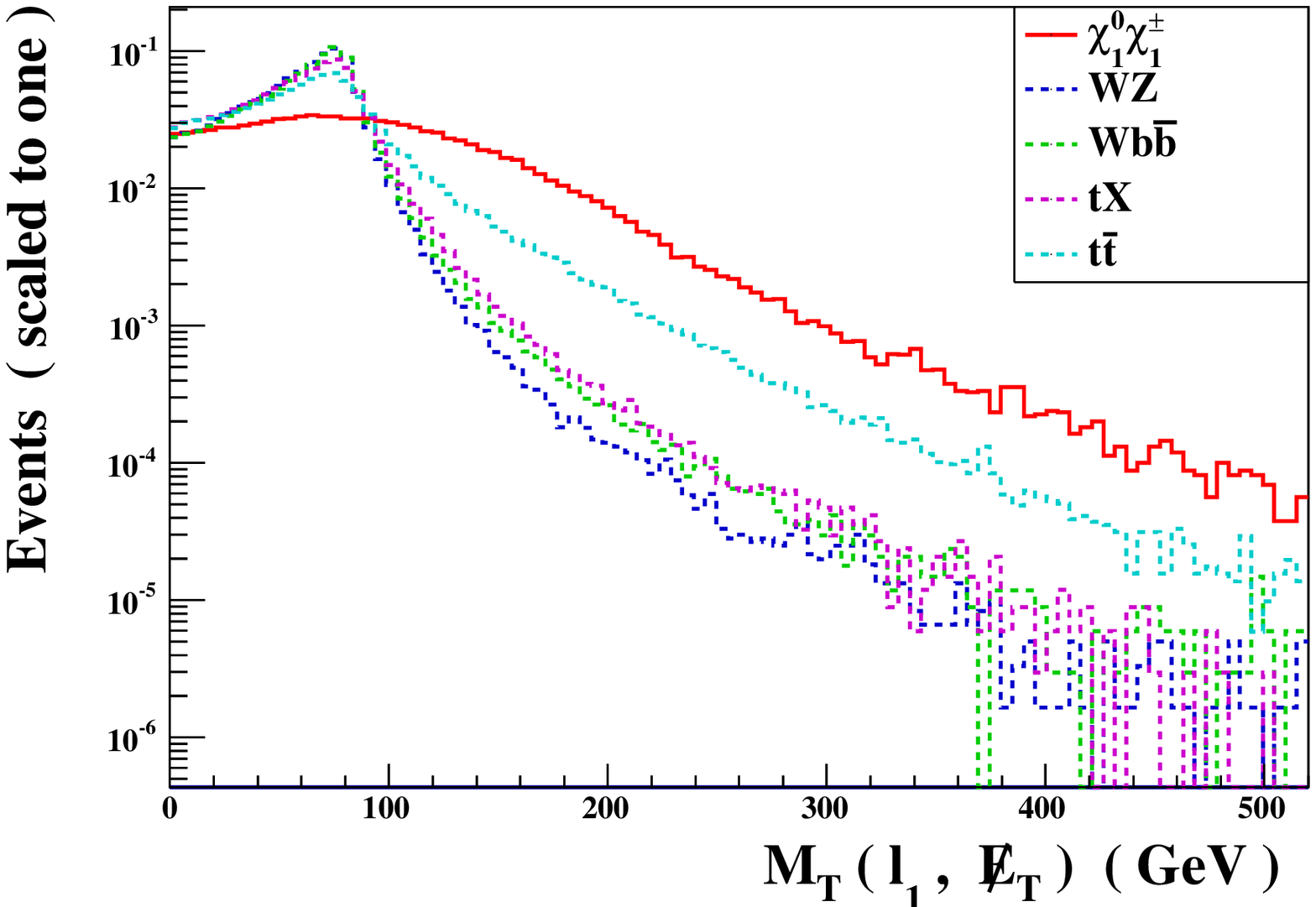}%
\includegraphics[scale=0.34]{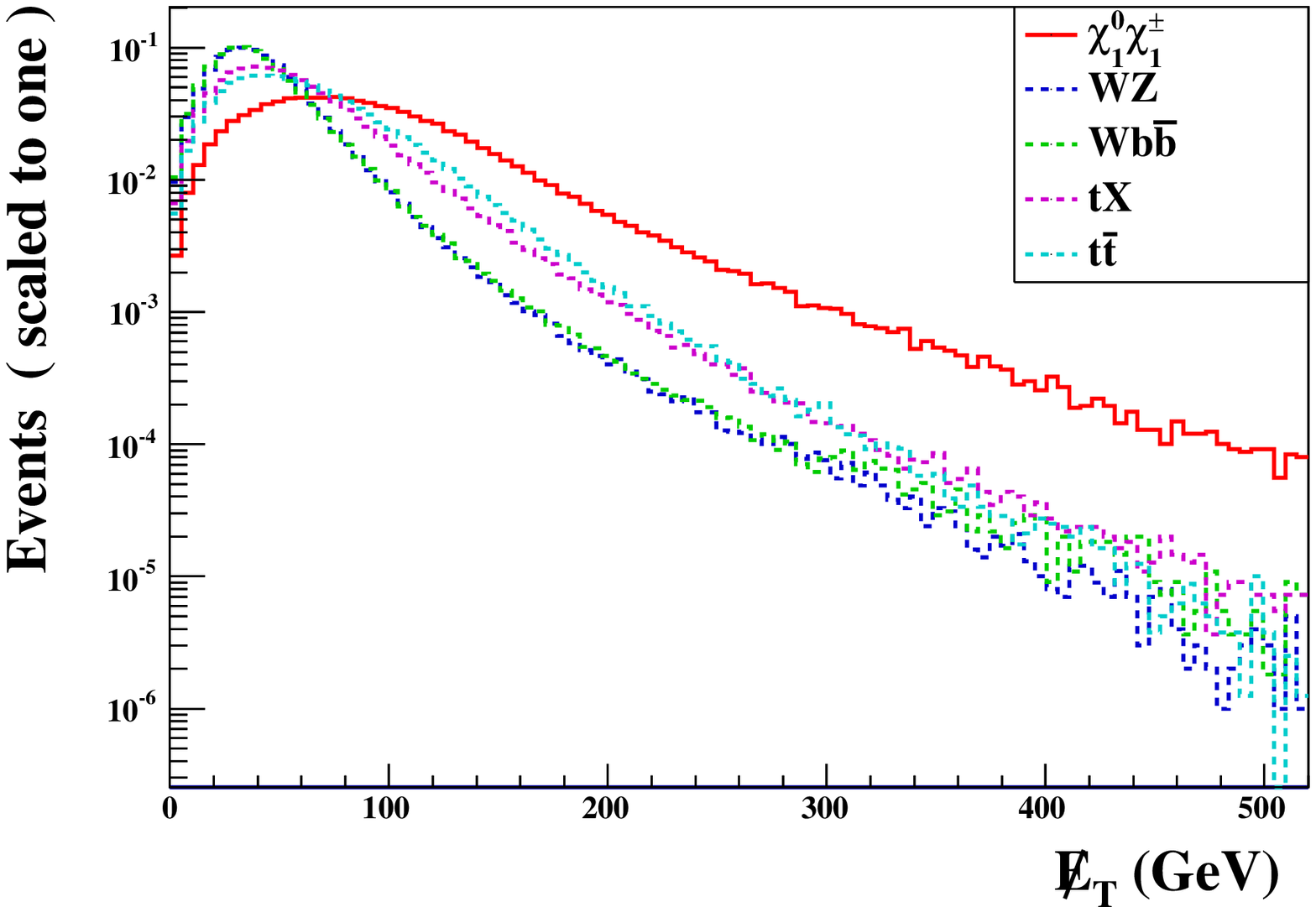}%
\hspace{0in}%
\caption{The normalized $M_{T}$ and $\missingET$ distributions for the signal
$pp\rightarrow\chi_{1}^{0}\chi_{1}^{\pm} \rightarrow h W^{\pm} G^{\prime}G^{\prime}
\rightarrow \ell^{\pm}  \nu b\bar{b} G^{\prime}G^{\prime}
\to \ell+2b + \missingET$ and
backgrounds at the LHC with $\sqrt{s}=14$~TeV. For the signal we fixed the relevant
mass parameter as $M_{2}=200$~GeV, $\mu=1.0$~TeV, $M_{1}=1.5$~TeV. The other parameters are
same as in Fig.~\ref{figetmt}.}
\label{figetmt2}
\end{figure}

In Fig.~\ref{figetmt2} we present the normalized $M_{T}$ and
$\missingET$ distributions of the signal and backgrounds at the 14
TeV LHC. It is expected that the peak of the transverse mass
distribution for the backgrounds with a single $W$ is around
$m_{W}$. Including di-leptonical channels, the shape of the curves
for top pair production should be a little different. We can observe
that the transverse mass cut should be effective for suppressing the
backgrounds. In the missing transverse energy distribution, we see
that the signal has a slightly harder $\missingET$ spectrum due to
the contribution of pseudo-goldstino. Thus a hard cut on $\missingET$ will
further reduce the backgrounds. At last we employ the following
selections for this signal:
\begin{itemize}
\item
Basic selection: one isolated lepton with $p_{T} > 40$ GeV, $\vert \eta \vert < 2.5$
and two tagged $b$-jets with $p_{T}^{b_{1},b_{2}} > 60,40$ GeV, $\vert \eta \vert < 2.5$.
\item
The invariant mass of $b$-jets must be within $\vert m_{bb}-m_{h} \vert < 25$ GeV.
\item $M_{T} > 100$ GeV.
\item $\missingET > 120$ GeV.
\end{itemize}

\begin{table}
\caption{The numbers of events for signal
$pp\rightarrow\chi_{1}^{0}\chi_{1}^{\pm} \rightarrow h W^{\pm} G^{\prime}G^{\prime}
\rightarrow \ell^{\pm} b\bar{b}  \nu G^{\prime}G^{\prime}
\to \ell+2b + \missingET$
and backgrounds for the LHC with $\sqrt{s} = 14$ TeV and 100 fb$^{-1}$
of integrated luminosity. } \vspace{0.2cm}
\begin{tabular}{||c||c|c|c|c||c||}
 \hline \hline
\cline{1-6}
 cut &~~$W Z\rightarrow \ell\nu b\bar{b}$~~  &  ~$W b \bar{b} \rightarrow \ell\nu b \bar{b} $~
&  ~$tX \rightarrow \ell b\nu X$~  & ~$t \bar{t} \rightarrow b \bar{b} \ell \nu \ell \nu(qq^{\prime})$~
& ~~signal~~ \\
 \hline
 basic selection &  373  &   7845  &  50015  &  796066  &  956  \\
 \hline
 $\vert m_{bb}-m_{h} \vert < 25$ GeV  &  82   &  1913   &   13164   &   199941   &   769  \\
 \hline
 $M_{T} > 100$ GeV  &  4  &  220  &  1215   &  27845  &  367   \\
 \hline
 $\missingET > 120$ GeV  &  1  &  3  &  69   &  2617  &  149   \\
 \hline
 \hline
\end{tabular}
\label{table3}
\end{table}

In Table \ref{table3} we display the cut flow for the signal and
backgrounds at the LHC with $\sqrt{s}=14$ TeV and an integrated
luminosity of 100 fb$^{-1}$. Note that we have normalized the
dominant  $t\bar{t}$ background to NNLO \cite{ttbnnlo}. We see that
the invariant mass cut strongly suppresses the backgrounds, while
having little effect on the signal. As we have shown in
Fig.~\ref{figetmt2}, the rather hard cuts on $M_{T}$ and $\missingET$
can efficiently reduce the SM backgrounds. We
observe from Table \ref{table3} that these cuts can almost remove
the $W b\bar{b}$ background. The dominant top pair and single top
backgrounds are also reduced by about several orders of magnitude.
However, the signal is only suppressed by a factor of seven.

\begin{table}
\caption{The number of the signal events
$pp\rightarrow\chi_{1}^{0}\chi_{1}^{\pm} \rightarrow h W^{\pm} G^{\prime}G^{\prime}
\rightarrow \ell^{\pm} b\bar{b}  \nu G^{\prime}G^{\prime}
\to \ell+2b + \missingET$
and its statistical significance for the LHC
with $\sqrt{s} = 14$ TeV and different luminosities. $S_{1}$ and $B_{1}$ stand for the
signal and background events after basic selection, while $S_{2}$ and $B_{2}$ stand for
the signal and background events after all the cuts.} \vspace{0.2cm}
\begin{tabular}{||c|c|c|c|c|c|c||}
 \hline \hline
 ~~~$\sqrt{s} = 14$ TeV~~~  &  ~100 fb$^{-1}$~  &  ~200 fb$^{-1}$~  &  ~300 fb$^{-1}$~  &  ~400 fb$^{-1}$~  &  ~500 fb$^{-1}$~  &  600 fb$^{-1}$  \\
\hline
 $S_{1{\rm [basic~selection]}}$  &  956  &  1912  &  2868  &  3824  &  4780  &  5736   \\
 \hline
 $S_{2{\rm [passing~all~cuts]}}$ &  149  &  298  &  447   &  596  &  745   &  894   \\
  \hline
 $S_{1}$/$\sqrt{ S_{1}+B_{1}}$  &  1.0  &  1.5  &  1.8  &  2.1  &  2.3  &  2.5   \\
 \hline
 $S_{2}$/$\sqrt{ S_{2}+B_{2}}$  &  2.8  &  4.0  &  4.8  &  5.6  &  6.3  &  6.8  \\
 \hline
 \hline
\end{tabular}
\label{table4}
\end{table}

In Table \ref{table4} we present the number of signal events and its
statistical significance for different luminosities at the 14 TeV
LHC. As expected, these optimization cuts improved the signal
significance efficiently. We see that the significance can reach
$5\sigma$ for an integrated luminosity of about 300 fb$^{-1}$. We
also notice that the ratio of signal-to-background is only about
$6\%$. This implies that the systematic uncertainty must be
controlled at percent level in order to detect the signal in this
case.

\subsection{Bino-like LOSP ($M_{1} < M_{2}, \mid\mu\mid$)}

In this case the lightest neutralino is bino-like
and its pair production cross section is small at the LHC
($10^{-6}$--$10^{-7}$ pb).
For the next lightest ordinary supersymmetric particle (NLOSP),
its components depend on the relative
values of $M_{2}$ and $\mu$. In the following we investigate the different
scenarios: (i) $\vert\mu\vert<M_{2}$, in which the next lightest neutrilino $\chi^{0}_{2}$
and chargino $\chi^{\pm}_{1}$ are higgsino-like;
(ii) $M_{2} < \vert\mu\vert$, in which the next lightest neutrilino $\chi^{0}_{2}$ and
chargino $\chi^{\pm}_{1}$ are wino-like.
In both scenarios, the leading production channels are the NLOSP pair production.
Since the decay of the neutral NLOSP is more sensitive to the SUSY parameters
than the charged NLOSP,
we therefore only explore the charged NLOSP pair ($\chi_{1}^{+}\chi_{1}^{-}$) production.
Here the chargino dominantly decays to a $W$ boson plus a bino-like LOSP $\chi_{1}^{0}$
or pseudo-goldstino $G^{\prime}$.

In case of a higgsino-like $\chi^{\pm}_{1}$, due to the relative large higgsino-bino mixing,
$\chi^{\pm}_{1}$ dominantly decays to $\chi^{0}_{1}$ and $W$ boson.
As discussed in Section II, a bino-like $\chi^{0}_{1}$ decays to
Higgs and pseudo-goldstino $G^{\prime}$.
Then this channel is  $p p \rightarrow \chi^{+}_{1}\chi^{-}_{1} \rightarrow \chi^{0}_{1} W^{+} \chi^{0}_{1} W^{-} \rightarrow h h W^{+} W^{-} G^{\prime} G^{\prime} $ (6.7 fb).
So its cross section is too small to be detected at the LHC.

In case of a wino-like $\chi^{\pm}_{1}$, there is little mixing
between bino and wino. Then $\chi^{\pm}_{1}$ will decay to pseudo-goldstino $G^{\prime}$ and
$W$ boson. Thus the signal is
\begin{equation}
p p \rightarrow \chi^{+}_{1}\chi^{-}_{1} \rightarrow W^{+} G^{\prime} W^{-} G^{\prime}
\rightarrow \ell^{+}\ell^{-} \nu\nu G^{\prime} G^{\prime}
\to 2\ell + \missingET
~~(l= e,\mu,\tau).
\end{equation}
The characteristic of this signal is two highly boosted leptons and
large missing transverse energy in the final state. This feature
will help to distinguish the signal from backgrounds. In our
analysis the bino-like LOSP neutralino is set as $M_{1}=200$ GeV.
Also, we set $M_{2}=500$ GeV and $\mu=1.0$ TeV, and other parameters
are the same as in the higgsino-like LOSP case.

The SM backgrounds come from the  di-boson productions of $WW$, $ZZ$
and $WZ$, the top pair and single top productions. The $WW$
background can be suppressed by requiring large missing transverse
energy. For $ZZ$ background process, when one of $Z$ bosons decays
to leptons and the other to neutrinos, it can resemble our signal.
These two leptons are different from the signal with highly boosted
leptons. Thus a high invariant mass cut on the two leptons could
reduce this background. For the $WZ$ background, it will fake the
signal only if one of three leptons in the final state is missing
detection. The two $W$ bosons produced in $t\bar{t}$ and $tW$
processes decay to leptons and thus can fake our signal. These
processes could be suppressed by applying $b$-jet and light jet
vetos. Since we require large transverse energy, the $W/Z$
production associated with a jet or photon will not be considered in
our work.

\begin{figure}[htbp]
\includegraphics[scale=0.35]{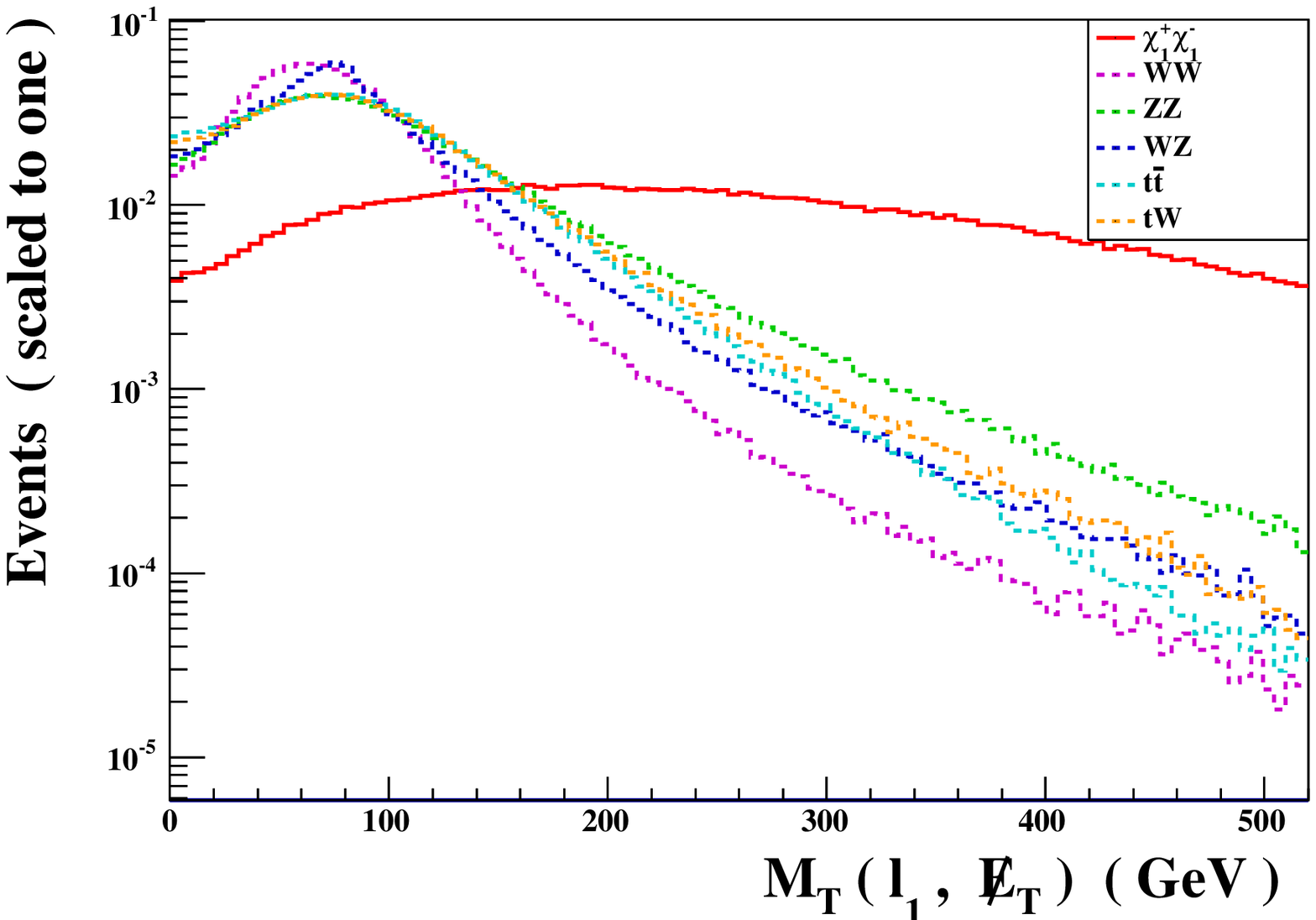}%
\includegraphics[scale=0.35]{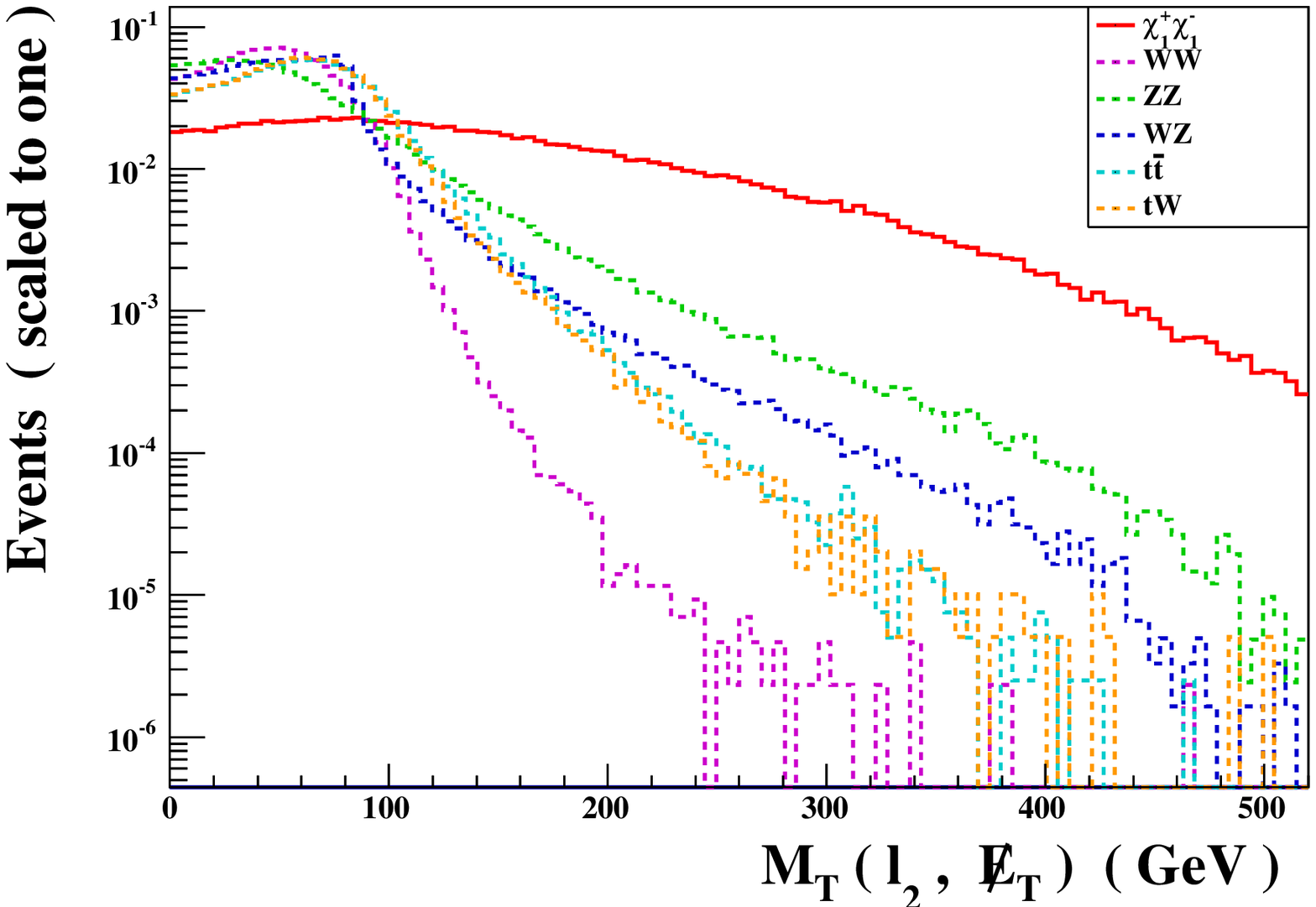}%
\hspace{0in}%
\includegraphics[scale=0.35]{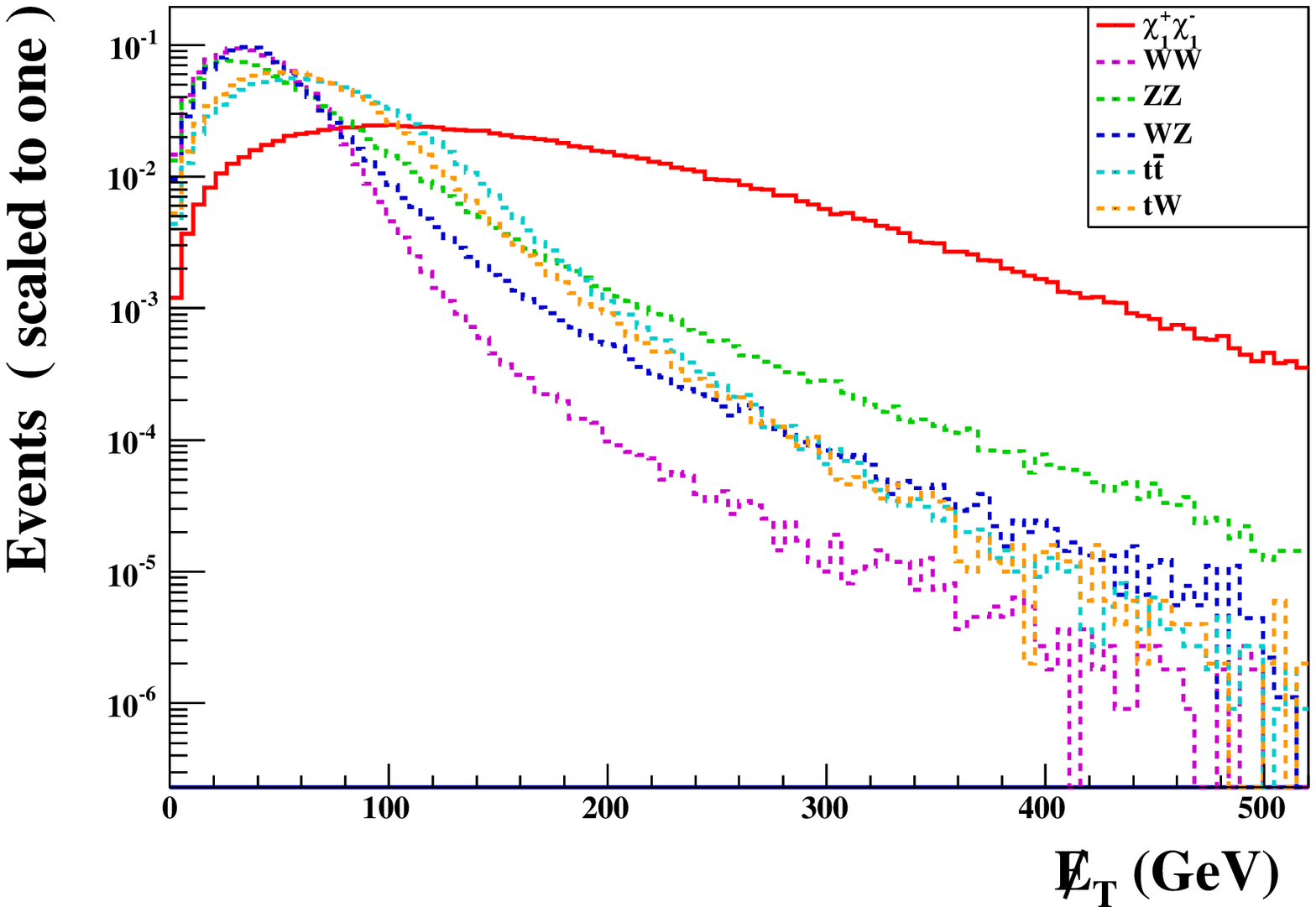}%
\caption{The normalized $M_{T}$ and $\missingET$ distribution for the signal
$p p \rightarrow \chi^{+}_{1}\chi^{-}_{1} \rightarrow W^{+} G^{\prime} W^{-} G^{\prime}
\rightarrow \ell^{+}\ell^{-} \nu\nu G^{\prime} G^{\prime}
\to 2\ell + \missingET$
and background processes at the LHC with $\sqrt{s}=14$~TeV. For the signal we fixed
the relevant mass parameters as  $M_{1}=200$~GeV, $M_{2}=500$~GeV, $\mu=1.0$~TeV. Other
parameters are same as in Fig.~\ref{figetmt}.} \label{figetmt3}
\end{figure}

In Fig.~\ref{figetmt3} we show the normalized $M_{T}$ distributions
of the hard and light charged leptons for the signal and backgrounds
at the 14 TeV LHC. Since both leptons in the signal come from the
decays of heavy particles, the signal has harder spectrum than
backgrounds in the $M_{T}$ distributions. We notice that the
backgrounds in the $M_{T}(\ell_{2}, \missingET)$
distribution have faster falling than in the $M_{T}(\ell_{1},
\missingET)$ distribution. Thus we will require a cut on
$M_{T}(\ell_{2}, \missingET)$ to suppress the
backgrounds. The normalized $\missingET$ distribution
for the signal and backgrounds is also presented in
Fig.~\ref{figetmt3}. We see the $\missingET$ distribution
for the signal is much harder than the signal due to extra pseudo-goldstino
contribution to the missing energy. We will apply a large missing
transverse energy cut to improve the signal significance. Based on
the above analysis, we apply the following selection for this
signal:
\begin{itemize}
\item
Basic selection: two opposite-sign leptons with $P_{T}^{\ell_{1},\ell_{2}} > 60,40$ GeV,
$\vert \eta \vert < 2.5$.
\item
$M_{T}(\ell_{2}, \missingET) > 120$ GeV.
\item
$\missingET > 120$ GeV.
\item
$M_{\ell^{+}\ell^{-}} > 140 $ GeV.
\item
Veto on tagged $b$-jets with $P_{T} > 20$ GeV and $\vert \eta \vert < 2.5$.
\item
Veto events with $P_{T}(j) > 50$ GeV and $\vert \eta \vert < 5.0$.
\end{itemize}

\begin{table}
\caption{The numbers of events for signal
$p p \rightarrow \chi^{+}_{1}\chi^{-}_{1}
\rightarrow W^{+} G^{\prime} W^{-} G^{\prime}
\rightarrow \ell^{+}\ell^{-} \nu\nu G^{\prime} G^{\prime}
\to 2\ell + \missingET$
and background at the LHC with $\sqrt{s} = 14$ TeV and 100 fb$^{-1}$
of integrated luminosity. } \vspace{0.2cm}
\begin{tabular}{||c||c|c|c|c|c||c||}
 \hline \hline
\cline{1-7}
 cut &  $W W\rightarrow \ell\ell\nu\nu$  & $Z Z\rightarrow \ell\ell\nu\nu$ &
 $W Z \rightarrow \ell\ell\ell\nu $
& $t \bar{t} \rightarrow b\bar{b}\ell\ell\nu\nu$ &
 $t W \rightarrow b\ell\ell\nu\nu$    & ~signal~ \\
 \hline
 basic selection &  30524  &   1524  &  1578  &  599505  &  52913 &  102 \\
 \hline
 $M_{T}(\ell_{2}, \missingET) > 120$ GeV  &  744   &  900   &   407   &   84647   &   6018 & 65.7 \\
 \hline
 $\missingET > 120$ GeV  &  12.6  &  582  &  180   &  14381  &  901 & 55.5  \\
 \hline
 $M_{\ell^{+}\ell^{-}} > 140 $ GeV   &  11.4  &  0.4  &  5.3   &  9759  &  643  & 43.3 \\
 \hline
 veto on b-jets  &  11.1  &  0.4  &  5.3  &  4107  &  334 & 43.1  \\
 \hline
 veto on light jet  &  6.1  &  0.3  &  1.9   &  124  & 17.9  &  37.3   \\
 \hline
 \hline
\end{tabular}
\label{table5}
\end{table}

\begin{table}
\caption{The number of events for the signal
 $p p \rightarrow \chi^{+}_{1}\chi^{-}_{1}
\rightarrow W^{+} G^{\prime} W^{-} G^{\prime}
\rightarrow \ell^{+}\ell^{-} \nu\nu G^{\prime} G^{\prime}
\to 2\ell + \missingET$
and its statistical significance for the LHC
with $\sqrt{s} = 14$ TeV and different luminosities. $S_{1}$ and $B_{1}$ stand for the signal
and background events after basic selection, while $S_{2}$ and $B_{2}$ stand for the signal and
background events after all the cuts.} \vspace{0.2cm}
\begin{tabular}{||c|c|c|c|c|c|c||}
 \hline \hline
 ~~~$\sqrt{s} = 14$ TeV~~~  &  ~100 fb$^{-1}$~  &  ~200 fb$^{-1}$~  &  ~300 fb$^{-1}$~  &  ~400 fb$^{-1}$~  &  ~500 fb$^{-1}$~  &  600 fb$^{-1}$  \\
\hline
 $S_{1{\rm [basic~selection]}}$  &  102  &  204  &  306  &  408  &  510  &  612   \\
 \hline
 $S_{2{\rm [passing~all~cuts]}}$ &  37.3  &  74.6  &  112   &  149  &  187   &  224   \\
  \hline
 $S_{1}$/$\sqrt{ S_{1}+B_{1}}$  &  0.12  &  0.17  &  0.21  &  0.25  &  0.28  &  0.30   \\
 \hline
 $S_{2}$/$\sqrt{ S_{2}+B_{2}}$  &  2.7  &  3.9  &  4.7  &  5.4  &  6.1  &  6.8  \\
 \hline
 \hline
\end{tabular}
\label{table6}
\end{table}

In Table \ref{table5}  we present the cut flow for the signal and
background events at the LHC with $\sqrt{s}=14$ TeV and an
integrated luminosity of 100 fb$^{-1}$. We have normalized the
dominant $t\bar{t}$  background to NNLO \cite{ttbnnlo}. We see that
the signal is overwhelmed by the backgrounds at the basic selection
level. As we expected, the $M_{T}$ cut on the light lepton can
suppress the backgrounds, while keeping most of the signal. This cut
is extremely effective for suppressing the $WW$ background. Then the
$WW$ background is further suppressed by a hard cut on $\missingET$.
The $WZ$ and $ZZ$ backgrounds with two leptons from
$Z$ decay are removed by requiring a large invariant mass of
leptons. The dominant reducible backgrounds $t\bar{t}$ and $tW$ are
suppressed strongly
 by the veto on $b$-jets and light jets.
After all cuts, the signal-to-background ratio is about $25\%$.

In Table \ref{table6} we display the number of signal events and its significance before and
after the cuts for different luminosities at the 14 TeV LHC. We see that the significance is
improved by these cuts efficiently. The significance can reach $5\sigma$ for a luminosity of
300--400 fb$^{-1}$.

\begin{table}
\caption{The statistical significances for three different cases at the LHC with $\sqrt{s}=8$ TeV and 21 $fb^{-1}$ } \vspace{0.2cm}
\begin{tabular}{||c|c|c|c|c||}
 \hline \hline
 ~~~$\sqrt{s} = 8$ TeV $\mathcal{L} = 21 fb^{-1}$ ~~~  &  $S$  &  $B$  &  $S/B$   &  $S/\sqrt{S+B}$    \\
\hline
 Higgsino-like LOSPs  &  11.2  &  90.1  &  0.12  &  1.11  \\
 \hline
 Wino-like LOSPs & 13.4  &  169.4  &  0.08  &  0.99   \\
  \hline
 Bino-like LOSPs  &  2.0  &  10.5  &  0.19  &  0.55 \\
 \hline
 \hline
\end{tabular}
\label{table7}
\end{table}

\vspace{0.5cm}

Finally, we note that the LHC searched the neutralinos and charginos with leptons plus missing $E_T$
at 7 and 8 TeV, and the observed events are in agreement with the SM backgrounds
(no excess), which gave some limits on the relevant parameter space \cite{EXP}.
 Since in our scenario the signals are quite rare compared with
the huge SM backgrounds (as shown in our results, only at 14 TeV LHC with a rather high luminosity
can our signals be possibly accessible), the current LHC limits at 7 and 8 TeV with rather
limited luminosities are not yet able to constrain the scenario under our consideration.
We numerically checked this and the results for the 8 TeV LHC are shown in Table \ref{table7}
(since the kinetic distributions of the signals and backgrounds for the 8 TeV LHC are similar to the results
for the 14 TeV LHC, we use the same cuts as for the 14 TeV LHC in each case).
We see that the  statistical significances are below 2$\sigma$ for a luminosity of  21 $fb^{-1}$.

\section{Conclusion}
Pseudo-goldstino is predicted in the multi-sector SUSY breaking scenario.
Comparing to the ordinary gravitino, it can couple to the visible sector
more strongly and
hence lead to some intriguing phenomenology at colliders.
In this scenario the lightest neutralino (chargino)
can decay into a pseudo-goldstino plus a $Z$-boson or Higgs boson ($W$-boson).
In this work we performed a Monte Carlo simulation for the direct productions of
the lightest neutralino and chargino followed by the decays to pseudo-goldstino.
Considering a higgsino-like, bino-like or wino-like lightest neutralino,
we found that the signal-to-background ratio ($S/B$) is
6\%--25\% and the statistical significance $S/\sqrt{S+B}$
is $5\sigma$ at the high luminosity LHC. So it is feasible to
explore such a multi-sector SUSY breaking scenario at the high luminosity LHC
if the background is known to percent level.

\section*{Acknowledgments}
Lin Wang acknowledges Prof.\ Johann H. K\"{u}hn and Prof.\ Matthias Steinhauser
for their warm hospitality. This work is supported by
by the Grant-in-Aid for Scientific Research (No.~24540246)
from Ministry of Education, Culture, Sports, Science and Technology (MEXT)
of Japan, by DFG through SFB/TR 9
``Computational Particle Physics'' and by the National Natural
Science Foundation of China under grant Nos. 11275245, 10821504 and
11135003.

\end{document}